\documentclass[twocolumn,showpacs]{revtex4}
\usepackage{amsfonts}
\usepackage{amsmath}
\usepackage{graphicx}
\begin{document}
\draft
\title{Measurement of the ac conductivity of a niobium single crystal in a swept
magnetic field}
\author{M.I. Tsindlekht, V.M. Genkin, G.I. Leviev, Y. Schlussel, and G. Masri}
\affiliation{The Racah Institute of Physics, The
Hebrew University of Jerusalem, 91904 Jerusalem,
Israel}

\begin{abstract}

We report results of experimental studies of ac susceptibility of Nb single crystal at low frequencies in swept magnetic fields applied parallel to the surface. Analysis of the experimental data shows that the swept magnetic field significantly changes the vortex conductivity in the bulk. It becomes dissipative with unexpected large frequency dispersion. At the surface, one observes a layer with enhanced conductivity in comparison to the bulk. This layer provides a considerably large contribution to the shielding and absorption of an ac field even in dc fields below $H_{c2}$. We have also demonstrated that the swept magnetic field apparently affects an ac response of the surface superconducting states.

\end{abstract}
\date{\today}
\maketitle

First measurements of the ac response of superconductors in a swept magnetic field were performed by M. Strongin \textit{et al.} in 1964~\cite{STR2}. This experiment demonstrated a qualitative difference between the results of point-by-point measurements and measurements in a swept field. Point-by-point measurements of the in-phase susceptibility showed full screening in magnetic field up to $H_{c3}$ and contrary to that in a swept field a partial penetration of the weak ac signal was detected for $H_{c1}<H_0<H_{c3}$.
Two years later Maxwell and Robbins~\cite{MAX} showed that in a mixed state the ac susceptibilities $\chi^{\prime}$ and $\chi^{\prime\prime}$ are a function of a single parameter $q=(dH_0/dt)/\omega h_0$, where $\omega$ is an excitation frequency, $h_0$ is the amplitude of the ac field and $dH_0/dt$ is the magnetic field sweep rate.
Schwartz and Maxwell~\cite{SCH} have calculated the ac susceptibility of a type-II superconductor in a swept magnetic field. They find that their model provides fair agreement with the experimental results of Ref.~\cite{MAX}. It should be noted that the model developed by Schwartz and Maxwell is essentially static ~\cite{SCH}.
A more comprehensive model of a low frequency response in a swept field for surface superconducting states (SSS)~\cite{PG}, $H_0>H_{c2}$, was developed by H. Fink ~\cite{FINK2}. Fink's model contains two parameters, $q$  and $p\equiv h_c/h_0$ which determine the ac response in a swept magnetic field. Here $h_c=4\pi J_c/c$, where $J_c(H_0)$ is the critical surface current. The effect of the swept magnetic field on the nonlinear response was investigated by authors of Ref.~\cite{CAMP}. It was found that the signal at the second harmonic appeared in a swept field while it was not observed in point-by-point measurements. In spite of all these investigations, the detailed physical picture of forming an ac response in a swept magnetic field is not clear.

In this paper, we report the first measurement of the bulk ac conductivity in the mixed state of type-II superconductor in a swept magnetic field. We demonstrate that at the surface there exists a layer of enhanced conductivity in comparison to the bulk. The conductivity of this layer has both dissipative and nondissipative components, while the bulk conductivity has actually dissipative component only. Substantial frequency dispersion was observed too. In high magnetic fields when the sample is in the SSS, i. e. for $H_{c2}<H_0<H_{c3}$, the swept magnetic field affects the ac response significantly.

 The ac response was measured by the pick-up coil method. The sample, with sizes 1 by 2.4 by 10 mm, was inserted into one of a balanced pair of coils, and the unbalanced signal was measured by a lock-in amplifier. The small unbalance signal of the empty coils was measured and substracted. RRR of this sample was $\approx 200$. The measurements were performed over a considerably wide range of frequencies (73 - 1465 Hz), sweep rates (SR) (5-30 Oe/sec), ac amplitudes, $h_0$, (0.02-0.2 Oe) as a function of the dc magnetic field. A "home-made" measurement cell of the experimental setup was adapted to a commercial SQUID magnetometer.  Both ac and dc fields were parallel to the longest sample axis. Magnetic field dependencies of the ac response were measured in point-by-point mode and in a swept magnetic field. For the measurements in a swept field, the standard power supply of the SQUID magnetometer solenoid was replaced by the external Oxford Instruments superconducting magnet power supply. A block-diagram of the experimental setup was published elsewhere~\cite{LEV2}.

\begin{figure}
\begin{center}
\leavevmode
\includegraphics[width=0.9\linewidth]{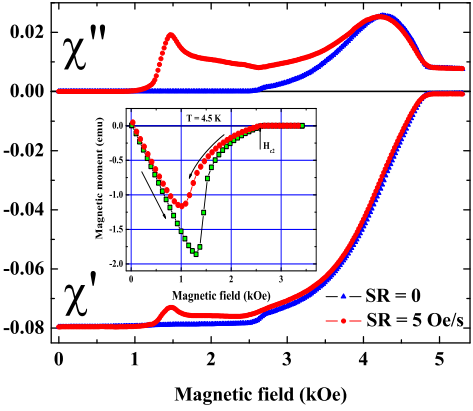}

\caption{(Color online)  Field dependencies of $\chi^{\prime}$ and $\chi^{\prime\prime}$ for SR 0 and 5 Oe/s at T = 4.5 K, frequency 146.5 Hz, and at $h_0=0.1$ Oe. Inset: Magnetization curve after ZFC.}
\label{f-1a}
\end{center}
\end{figure}

DC magnetization curve data (inset to Fig.~\ref{f-1a}) show that at T = 4.5 K H$_{c1}\approx 1$ kOe and H$_{c2}\approx 2.6$ kOe. The magnetization curve in the vicinity of H$_{c2}$ is roughly a linear function of H$_0$ and at this temperature the Ginzburg-Landau parameter $\kappa\approx 1.6$. The apparent normal conductivity is $\sigma\approx 1.1\times 10^{19}$ s$^{-1}$ as was determined from the ac response in magnetic field $H_0>H_{c3}$.

Fig.~\ref{f-1a} demonstrates the field dependence of $\chi^{\prime}$ and $\chi^{\prime\prime}$ in the point-by-point mode and in the swept field with SR 5 Oe/s and $h_0=0.1$ Oe. The swept field affects the ac response for $H_{c1}\leq H_0\leq H_{c3} $. So, the swept field affects the conductivities in a mixed and in the surface superconducting states.

\begin{figure}
\begin{center}
\leavevmode
\includegraphics[width=0.9\linewidth]{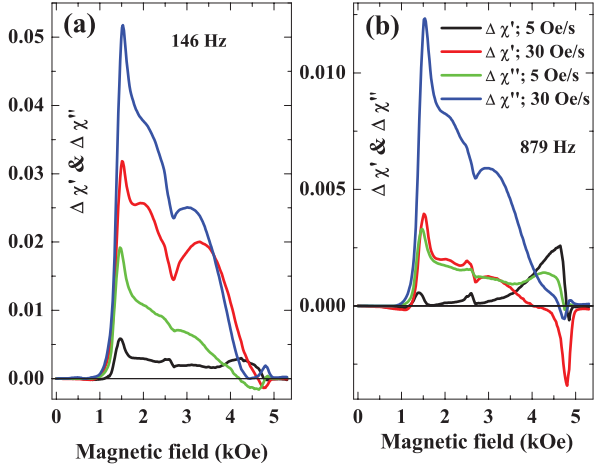}

\caption{(Color online)  Field dependence of $\Delta\chi=\chi_{sw}-\chi_{st}$ at 146.5 Hz (panel (a)) and 879 Hz (panel (b)), and at $h_0=0.1$ Oe. Symbols are the same in both panels.}
\label{f-1}
\end{center}
\end{figure}

Figures~\ref{f-1}a and~\ref{f-1}b present in detail the frequency dispersion in a swept magnetic field.
Here we show the difference between the susceptibility in a swept magnetic field, $\chi_{sw}$, and in a stationary field, $\chi_{st}$,  ($\Delta\chi=\chi_{sw}-\chi_{st}$) at two frequencies 146 Hz (panel (a)) and 879 Hz (panel(b)), at SR 5 Oe/s and 30 Oe/s as a function of $H_0$. It is clear that changes in $\chi$ due to the swept field is much stronger for low frequencies.

\begin{figure}
\begin{center}
\leavevmode
\includegraphics[width=0.9\linewidth]{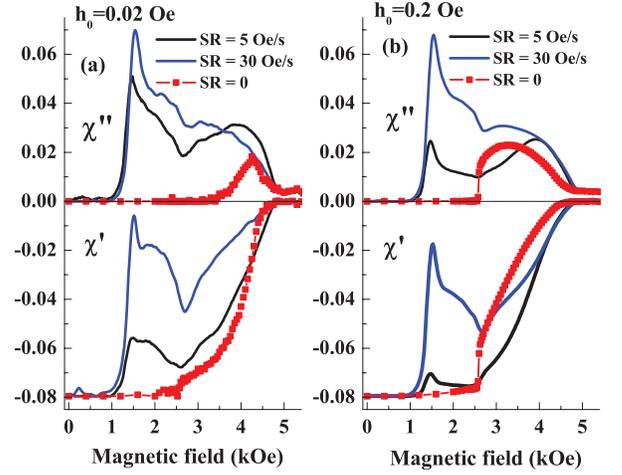}

\caption{(Color online)  Field dependence of $\chi$ at different SR and ac amplitudes for 73 Hz. Panel (a) - $h_0=0.02$ Oe; panel (b) - $h_0=0.2$ Oe.}
\label{f-2}
\end{center}
\end{figure}

Fig.~\ref{f-2} shows the ac response at two amplitudes of the ac field. Here we observe the qualitative difference between behavior at $H_{c1}<H_0<H_{c2}$ and $H_{c2}<H_0<H_{c3.}$. In the mixed state, the increasing of the ac amplitude does not change the character of the response, the swept field leads to a decreasing of the shielding and an increase the dissipation. In the area of surface superconductivity $H_0>H_{c2}$ it takes place only for weak ac amplitudes. At larger amplitudes, the swept field increases the shielding while the losses could be decreased.

The observed ac response is formed by both surface and bulk currents. For $H_0 > H_{c2}$ the interior of the sample is in a normal state with known conductivity and we obtain the surface current $J_s$ directly from the measured susceptibility. Assuming that the thickness of the surface layer is small in comparison to the sample sizes, we can write
$(1+4\pi\chi)h_0=(1+4\pi\chi_{\infty})(h_0-4\pi J_s/c)$,
where $\chi_{\infty}$ is the susceptibility for $H_0>H_{c3}$.
We obtain
\begin{equation}\label{Eq3}
J_s/h_0=(\chi_{\infty}-\chi)/(\chi_{\infty}+1/4\pi).
\end{equation}
In this procedure we actually substracted the contribution of the normal interior of the sample to the measured ac susceptibility.

\begin{figure}
\begin{center}
\leavevmode
\includegraphics[width=0.9\linewidth]{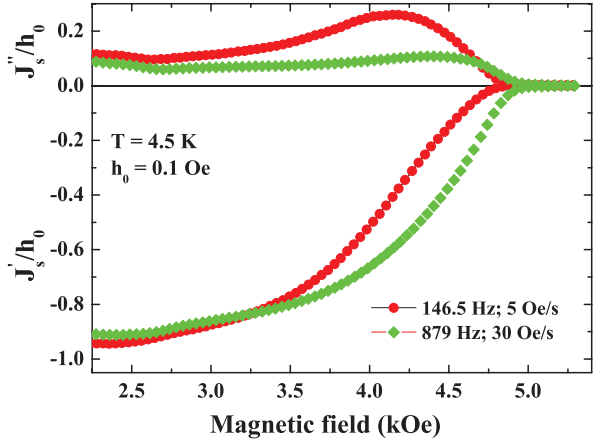}

\caption{(Color online)
  $J_s/h_0$ as a dc magnetic field function in the SSS. Parameters $q$  and $p$ coincide for these curves.}
\label{f-7a}
\end{center}
\end{figure}

We used Eq.~(\ref{Eq3}) for testing Fink's result~\cite{FINK2} that the ac response in the SSS in a swept field is a function of only two parameters $q$ and $p$. To this end, we compared the data for two different sets of frequencies and sweep rates at constant amplitude of excitation, so that  parameters $q$  and $p$ were the same for each set.  Fig.~\ref{f-7a} presents the obtained  $J_s/h_0$ as a function od dc field at  frequency 146.5 Hz,  SR 5 Oe/s and at frequency 879 Hz, SR 30 Oe/s and ac amplitude $h_0=0.1$ Oe.  For these two sets of parameters $q$  and $p$ are equal for any given magnetic field.  Experiment clearly shows that these two parameters alone do not provide an adequate description of the experimental data.

For $H_{c1}<H_0 < H_{c2}$ the problem of separation of surface and bulk ac currents from the experimental data is more complicated. The ac susceptibility of the sample in a longitudinal ac magnetic field could be found, assuming that bulk conductivity is $\sigma_{eff}=\sigma_1+i\sigma_2$, from
\begin{equation}\label{Eq1}
\frac{\partial^2h_z}{\partial x^2}+\frac{\partial^2h_z}{\partial y^2}-\left (\frac{1}{\lambda^2}-\frac{2i}{\delta^2}\right ) h_z=0.
\end{equation}
Here $\delta=c/\sqrt{2\pi\sigma_1\omega}$, and $\lambda=c/\sqrt{2\pi\sigma_2\omega}$. In this equation we neglect a small demagnetization factor assuming that the sample size in the field direction (axis Z), is much greater than its sizes $2L_x$, $2L_y$ in the plane $z=0$. Straightforward solution of this equation with boundary conditions $h_z=\alpha h_0$ gives
\begin{equation}\label{Eq2}
\chi=(\alpha Z_1+\alpha Z_2-1)/4\pi,
\end{equation}
 where
 $$Z_1=8\sum_{m=1,3...} \tanh(k_mL_y)/\pi^2m^2k_mL_y;$$ $$Z_2=8\sum_{m=1,3...} \tanh(q_mL_x)/\pi^2m^2k_mL_x;$$ $k_m^2=(\pi m/2L_x)^2+k^2$, $q_m^2=(\pi m/2L_y)^2+k^2$, $k^2=-4\pi i\omega\sigma_{eff}/c^2=1/\lambda^2-2i/\delta^2$. The phenomenological complex parameter $\alpha$ takes into account the possible existence of the surface layer with enhanced conductivity. This layer provides a jump of the ac field at the boundary because of the surface current. For example, in the model with thin layer of thickness \textit{d} at the surface, with conductivity $|\sigma_s|\gg |\sigma_{eff}|$ $\alpha=1/(1+k_s^2d/k)$, where $k_s^2=-4\pi i\omega\sigma_s/c^2$.

\begin{figure}%
\begin{center}
\leavevmode
\includegraphics[width=0.9\linewidth]{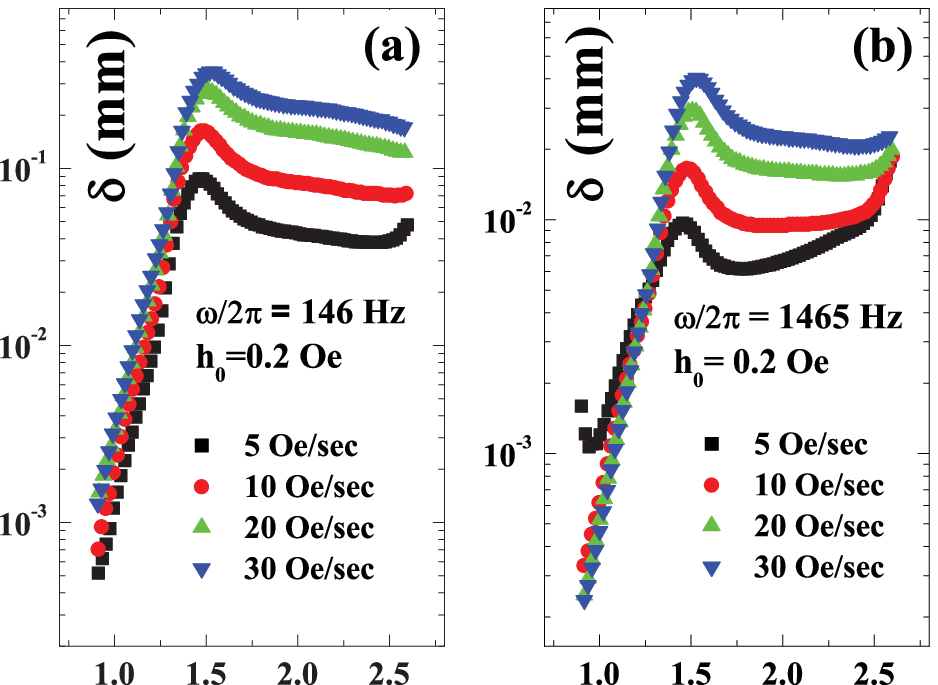}
\includegraphics[width=0.9\linewidth]{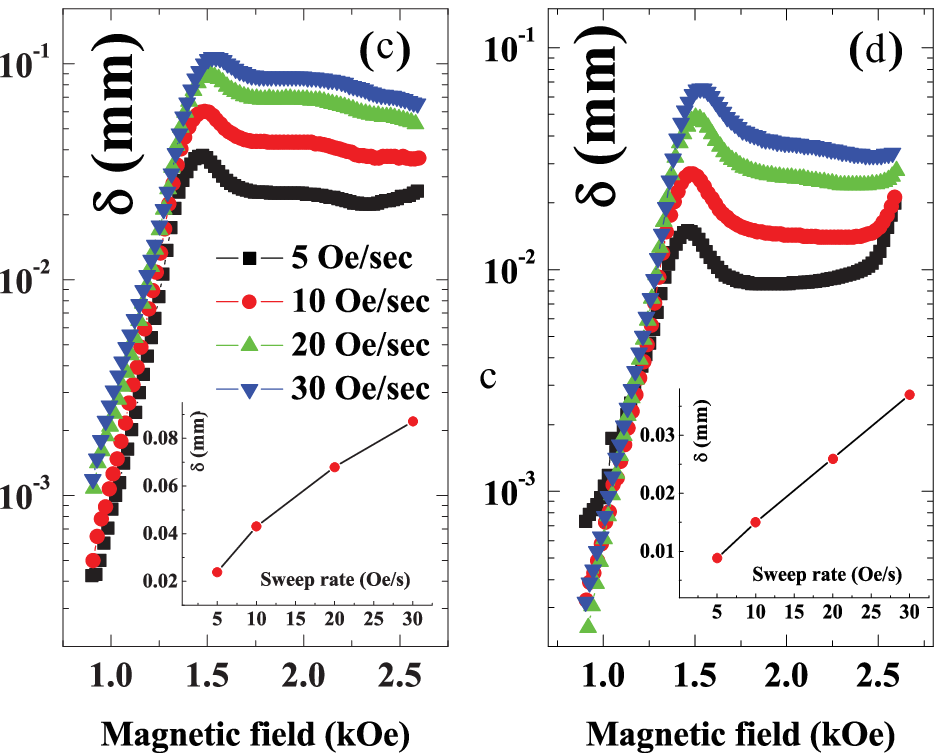}

\caption{(Color online)  Field dependence of $\delta$ at frequencies 146 Hz - panel (a) and 1465 Hz - panel (b) for different sweep rates. Field dependence of $\delta$ at frequency 879 Hz, different sweep rates and excitation amplitudes. Panel (c) - $h_0=0.02$ Oe  and panel (d) - 0.2 Oe. Insets to both panels: Sweep rate dependence of $\delta$ at $H_0=2$ kOe. Symbols are the same for (c) and (d) panels.}
\label{f-3}
\end{center}
\end{figure}

As far as the point-by-point data are concerned, there is complete shielding for $H_0<H_{c2}$ and we could conclude only that the penetration depth is very small. In the swept field, we observe an ac response, which differs from complete shielding, and this could be used for extracting the bulk conductivity from experimental data. We found that it is impossible to map adequately the observed value $\chi$ in the swept field onto the $\sigma_{eff}$ using Eq.~(\ref{Eq2}) if we neglect the surface current, i.e. assume that $\alpha  = 1$.
In general, the two unknown quantities $\sigma_{eff}$ and $\alpha$ could not be found using only $\chi$ data without any additional conditions. We assumed that the total surface current has a minimal possible value because it localized in a thin surface layer. This condition permits one to find both the bulk $\sigma_{eff}$ and $\alpha$, (i. e. the value of the surface current $J_s\propto 1-\alpha$).
 Results of this approach for $\delta$ are shown in Figs.~\ref{f-3}.
Bulk conductivity has a dissipative character and $\sigma_{2}$ is very small in comparison to $\sigma_{1}$.
Approximately exponential growth of $\delta$ with the dc field for 1 kOe$< H_0<$1.5 kOe is evidently due to a fast increase of the vortex density in the bulk. With increasing sweep rate the conductivity decreases and the effective penetration depth, $\delta$, increases. There is considerable frequency dispersion. Increasing the frequency from 146.5 Hz to 1465 Hz leads to an approximate 10 fold decrease of $\delta$, i.e. the conductivity increased linear with frequency. $\delta$ depends also on the excitation amplitude, Fig.~\ref{f-3}(c,d).
The sweep rate dependence of $\delta$ for $H_0=2$ kOe, $h_0=0.02$ Oe and 0.2 Oe is presented in insets to Fig.~\ref{f-3}(c,d). This dependence is almost linear at low sweep rates.

\begin{figure}
\begin{center}
\leavevmode
\includegraphics[width=0.9\linewidth]{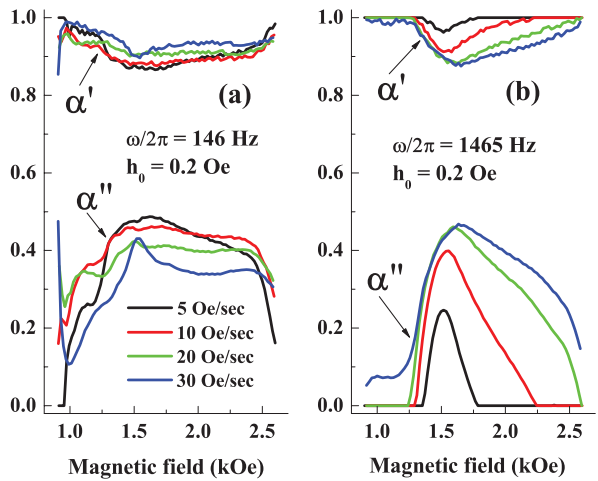}

\caption{(Color online)  Field dependencies of surface current, parameter $\alpha$, for $H_{c1}<H_0<H_{c2}$  at frequency 146.5 Hz (panel (a)) and 1465 Hz (panel (b)) for different sweep rates. Excitation amplitude  $h_0=0.2$ Oe. Symbols on both panels are the same.}

\label{f-6}
\end{center}
\end{figure}

At the surface there is an ac current which has both dissipative and nondissipative components. The value of the surface current is not negligibly small and, in general, the dissipative component is larger than the nondissipative one. Fig.~\ref{f-6} demonstrates the field dependencies of the parameter $\alpha$ for different frequencies and sweep rates. The frequency dispersion of $\alpha$ is rather small.

The field dependence of $\chi$ is different for mixed and surface superconducting states.
 This is clear as seen from Fig.~\ref{f-2} where we observe a minimum in $\chi^{\prime\prime}$ at $H_{c2}$. So that the mechanisms responsible for the ac response in the mixed state and in the surface superconducting states are different.
We found that at $H_{c1}<H_0<H_{c2}$ conductivity in the interior of the sample becomes dissipative in a swept field and $\delta$ is considerably smaller than the normal skin depth.
It seems plausible to relate the obtained $\delta$ with the Campbell penetration depth, $\lambda_C$, i.e. with penetration due to vortex lattice deformations~\cite{CAMPB}. But $\delta$ depends on frequency while $\lambda_C$ is actually frequency independent. The observed large $\delta$ could be ascribed to the appearance of weakly pinned vortices in the bulk due to a dc current that exists in the sample in a ramped dc field. These vortices have increased mobility and increase the penetration of the ac field in the sample.
It is known that in dc field parallel to the surface  there is a surface layer with low vortex density if the magnetic susceptibility of the sample differs from $-1/4\pi$~\cite{TERN}. In addition, at the surface, the order parameter could be larger then in the bulk~\cite{FINK1}. As a result, the conductivity of these layers could have an enhanced value in comparison to the bulk. In any case, our experimental data show that in a swept magnetic field we can observe an existence of the surface layer with enhanced conductivity  and this conductivity has both considerably large dissipative and nondissipative components.

For large dc fields, $H_{c2}<H_0<H_{c3}$, the experimental picture is better defined.
In these fields, the bulk of the sample is in a normal state and we have to deal with a conducting surface layer. The surface current is defined from the experimental data using Eq.~(\ref{Eq3}) without any additional assumptions. We noted that the swept field also considerably changes the ac response of the SSS, Fig.~\ref{f-1a}. The swept field increases losses at the surface and simultaneously decreases the shielding current. Effect of the swept field on the ac response reflects the dynamics of the surface superconducting states.

In conclusion, we have presented for the first time the results of measuring ac vortex conductivity in a swept magnetic field. We showed that for wide range of dc fields, $H_{c1} < H_0 < H_{c3}$, the surface layer with enhanced conductivity exists and provides a considerable contribution to the ac response in a swept magnetic field. Existence of this layer has to be taken into account in any treatment of the ac response of the superconductors. For $ H_0 > H_{c2}$ the existence of this layer is an unambiguous manifestation of the surface superconductivity while for $ H_0 < H_{c2}$ the existence of such a layer is the most intriguing part of our observations.

The authors are deeply grateful to I. Felner for
useful discussions. One of the authors (M.I.T.) wishes to thank B. Pla\c{c}ais for providing high quality single crystal Nb. This work was supported by the Israeli Ministry of Science (Israel-Russia fund), and by the Klatchky foundation for superconductivity.

\end{document}